\documentclass[12pt,a4paper]{article}
\usepackage{graphicx,amsmath,amssymb,cite,url}

\newcommand{\sla}[1]{\ifmmode%
  \setbox0=\hbox{$#1$}%
  \setbox1=\hbox to\wd0{\hss$/$\hss}\else%
  \setbox0=\hbox{#1}%
  \setbox1=\hbox to\wd0{\hss/\hss}\fi%
  #1\hskip-\wd0\box1 }

\begin{document}

\begin{titlepage}
\begin{center}
{\Large \bf Inadequacy of zero-width approximation for a light Higgs boson signal}\\[1.2cm]
Nikolas Kauer\\[0.3cm]
\textit{Department of Physics, Royal Holloway, University of London, 
Egham Hill, 
Egham TW20 0EX, 
United Kingdom}\\[1.5cm]
\textbf{Abstract}
\end{center}
\begin{quote}
The zero-width approximation (ZWA) restricts the intermediate unstable 
particle state to the mass shell and, when combined with the decorrelation 
approximation, fully factorizes the production and decay 
of unstable particles.  The ZWA uncertainty is expected to be of  
${\cal O}(\Gamma/M)$, where $M$ and $\Gamma$ are the mass and width of the 
unstable particle.  We review the ZWA and demonstrate that errors can be much 
larger than expected if a significant modification of the Breit-Wigner lineshape 
occurs.
A thorough examination of the recently discovered candidate 
Standard Model Higgs boson is in progress.  For $M_H\approx 125$\ GeV, one has 
$\Gamma_H/M_H < 10^{-4}$, which suggests an excellent accuracy 
of the ZWA.  We show that this is not 
always the case.  The inclusion of off-shell contributions is essential 
to obtain an accurate Higgs signal normalization at the $1\%$ precision level. 
For $gg\to H \to VV$, $V= W,Z$, 
\mbox{${\cal O}(5\,$--$\,10\%)$} corrections occur due to an enhanced Higgs 
signal in the 
region $M_{VV} > 2M_V$,
where also sizable Higgs-continuum interference occurs. 
We discuss how experimental selection 
cuts can be used to suppress this region in search channels where the 
Higgs mass cannot be reconstructed.
We note that $H\to VV$ decay modes in non-gluon-fusion channels 
are similarly affected. 
\end{quote}

\vspace{.3cm}


\textit{email}: n.kauer@rhul.ac.uk
\end{titlepage}

\section{Introduction \label{sec:intro}}

The theoretical study of unstable particles has a long history.  Initial work 
focused on an adequate field-theoretical definition of the mass and lifetime of an 
unstable particle \cite{Matthews:1958sc,Matthews:1959sy,Jacob:1961zz}.
At the time, a mass and width definition based on the position of the 
complex pole of the unstable particle propagator was also 
proposed \cite{complexpole1,complexpole2,complexpole3}.
A consistent quantum-field-theoretical formalism for the description 
of unstable particles with unitary, renormalizable and causal $S$-matrix was 
developed in Ref.~\cite{Veltman:1963th}.  Extending previous work to gauge 
theories, Refs.~\cite{Stuart:1991xk,Sirlin:1991fd} proposed a perturbative 
expansion of the exact scattering amplitude about the complex 
pole of the unstable 
particle propagator and a subsequent consistent expansion in powers of the 
coupling constant.  Thus obtaining an order by order gauge-invariant definition of 
the scattering amplitude in perturbation theory.  The 
gauge-invariant \cite{Gambino:1999ai} mass and width of the unstable particle in 
the corresponding pole scheme differ from the quantities in the commonly used 
on-shell scheme (see also Refs.~\cite{Willenbrock:1991hu,Kniehl:1998fn}), 
which exhibit a gauge dependence in higher orders.\footnote{%
The gauge-independent definition of partial widths and branching ratios has been 
studied in Refs.~\cite{Grassi:2000dz,Grassi:2001bz}.}  
The generalization of the complex pole approach to one-loop calculations with 
multiple, charged unstable particles has been discussed in 
Ref.~\cite{Aeppli:1993rs}.

In Refs.~\cite{Beneke:2003xh,Beneke:2004km,Chapovsky:2001zt,Falgari:2010sf,Falgari:2013gwa}, the complex pole approach has been generalized using the effective 
field theory formalism.  The effective theory is constructed by systematically 
expanding cross sections in powers of two parameters: the coupling constant 
($\alpha$) and a measure for the off-shellness of the unstable particle propagator 
($\delta$), which is of order $\Gamma/M$ in the region of validity, i.e.\  
close to resonance.\footnote{In calculations, applying the method of 
regions \cite{Beneke:1997zp,Jantzen:2011nz} can be more convenient than the
explicit procedure.}  The effective field theory approach yields a gauge-invariant 
expansion that can in principle be truncated at arbitrarily high orders in $\alpha$ 
and $\delta$ and that permits an extension, for instance, to threshold regions.  

An alternative approach that works more automatically and is expected to give 
precise numerical results in all kinematic situations, but is theoretically 
less rigorous, is the complex-mass 
scheme \cite{Denner:1999gp,Denner:2005fg,Denner:2006ic}.
At tree level, 
it exploits the fact that the full amplitude (without self-energy resummation)
is gauge independent and remains so if the mass of the unstable particle, $M$,
is systematically replaced with $\sqrt{M^2-iM\Gamma}$, thus including the 
higher-order self-energy corrections that remove the real pole of the unstable 
particle propagator.  Note that this procedure results, for instance, in a 
complex weak mixing angle: $\cos\theta_W=M_W/M_Z$.
The introduced spurious terms are corrections of 
${\cal O}(\Gamma/M)={\cal O}(\alpha)$ in the resonant as well as nonresonant 
phase space regions. 
Beyond tree level, the complex-mass scheme is implemented by splitting the 
real bare mass into a complex renormalized mass and a complex counterterm 
in the Lagrangian.  The complete renormalization prescription is given 
in Ref.~\cite{Denner:2005fg}.  
The complex-mass scheme has primarily been employed in multi-leg calculations 
at tree and one-loop level,\footnote{Note that one-loop integrals 
with complex internal masses are required.} 
but can in principle be extended to higher orders.
The theoretical weakness of the scheme is that the ad hoc 
introduction of complex masses violates unitarity.  It 
is plausible that the unitarity-violating terms are of higher 
order, but a rigorous proof is still outstanding.

The outlined evolution of methods for perturbative calculations 
that involve unstable particles illustrates that despite the
impressive progress no formalism has been developed yet that on 
the one hand has a rigorous field-theoretical foundation and on 
the other hand provides a practicable and efficient implementation 
which returns reliable results of the desired precision for all 
phenomenologically relevant observables.
In this light, it is suggestive to revisit approximations 
where unstable particle states do not feature a continuous 
invariant mass spectrum, but are instead on-mass-shell.

The simplest such approximation is to treat unstable particles
as stable external particles.  The stable-particle approximation 
can be used 
to obtain results for inclusive observables with an expected 
uncertainty of ${\cal O}(\Gamma/M)$.  But, it does not allow
to calculate the differential cross section for processes that 
include the decay of the unstable particle.  This is, however,
essential for collider phenomenology, because typically only the 
decay products can be detected.  Furthermore, as shown in 
Ref.~\cite{Veltman:1963th}, in a sound perturbative field theory 
unstable particles do not occur as external states.\footnote{The absence 
of asymptotic final states for unstable particles is plausible.}

The zero-width approximation (ZWA) \cite{Pilkuhn}, 
a.k.a.\ narrow-width approximation, 
is a well-known on-shell approximation that is not affected 
by these shortcomings.  It exploits the asymptotic equality 
of the squared modulus of the unstable (scalar) particle 
propagator with 4-momentum $q$ to $\pi/(M\Gamma)\,\delta(q^2-M^2)$ 
in the limit $\Gamma\to 0$.  The 
Dirac delta function restricts the unstable particle to on-shell
states without otherwise affecting the production and decay
subprocesses.  In general, an uncertainty of ${\cal O}(\Gamma/M)$ is 
expected for the ZWA.  For many known and predicted unstable particles 
one finds that $\Gamma/M$ is of ${\cal O}(1\%)$, which implies
a ZWA error that is similar to other theoretical and experimental 
errors in collider experiments.  The ZWA is therefore widely applicable.
An exception are heavy Higgs bosons ($M_H \gtrsim 300$ GeV): 
for a Standard Model (SM) Higgs boson with mass of $300$ GeV, 
for instance, the expected ZWA uncertainty is of ${\cal O}(3\%)$ and 
increases to ${\cal O}(20\%)$ for $M_H = 600$ GeV.  The ZWA 
is evidently not adequate for a heavy Higgs boson, which rather 
requires a detailed description of the lineshape (Higgs invariant 
mass distribution), which has been developed recently in 
Refs.~\cite{Goria:2011wa,Franzosi:2012nk}.
On the other hand, for the recently discovered \cite{Aad:2012tfa,Chatrchyan:2012ufa}
Higgs boson with $M_H\approx 125$~GeV, one has
$\Gamma_H/M_H < 10^{-4}$ (in the SM), which suggests an 
outstanding accuracy of the ZWA.  However, as shown in 
Ref.~\cite{Kauer:2012hd} for inclusive cross sections and 
cross sections with experimental selection cuts, the ZWA is 
in general not adequate and the error estimate 
${\cal O}(\Gamma_H/M_H)$ is not reliable for a light Higgs boson.

The review is organized as follows: 
In Sec.~\ref{sec:zwa}, a general discussion of the ZWA and its 
uncertainty is presented.  Then, the ZWA inadequacy for a light 
Higgs boson signal is elucidated in Sec.~\ref{lhzwa}, followed by 
a summary in Sec.~\ref{summary}.


\section{Zero-Width Approximation: Definition and Uncertainty \label{sec:zwa}}


Consider the squared modulus of the unstable scalar particle propagator with 
4-momentum $q$, mass $M$ and width $\Gamma$,
\begin{equation}
D(q^2,M,\Gamma) = \frac{1}{\left(q^2-M^2\right)^2+(M\Gamma)^2}\,,
\label{eq:BW}
\end{equation}
where the fixed-width Breit-Wigner scheme has been used.  
We note that the field-theoretical definition of the 
on-shell mass $M$ and on-shell width $\Gamma$ of an unstable particle are not fully 
satisfactory and that the complex-pole scheme is theoretically better 
motivated (see Sec.~\ref{sec:intro}).  However, the propagator in the 
complex-pole scheme, $1/(q^2-s_{\mathrm{pole}})$ with 
$s_{\mathrm{pole}}=\mu^2 - i\mu\gamma$ (see e.g.\ Ref.~\cite{Goria:2011wa}), 
can be obtained to very high precision by substituting $M\to \mu$ and 
$\Gamma\to\gamma$ in Eq.~(\ref{eq:BW}).\footnote{%
The numerical difference between $D(q^2,\mu,\gamma)$ and the exact 
complex-pole scheme result is negligible in single precision.}  
The scheme choice does therefore not affect our discussion of the ZWA.
The formal definition of the ZWA is based on the following expansion:
\begin{equation}
D(q^2,M,\Gamma) =
C\,\delta\!\left( q^2 - M^2\right) 
+ PV\,\Bigl[ \frac{1}{\left( q^2 - M^2\right)^2}\Bigr] + \sum_{n=0}^N\,c_n(\alpha)\,
\delta_n\!\left( q^2 - M^2\right),
\label{BWexp}
\end{equation}
where $PV$ denotes the principal value (understood as a distribution),
$\delta_n(x)$ is related to the $n$th derivative of the delta function by
$\delta_n(x)= (-1)^n/n!\,\delta^{(n)}(x)$, and the expansion is in terms of 
the coupling constant $\alpha$, to a given order $N$.  The prefactor of 
the delta function is given by
\begin{equation}
C= \frac{\pi}{M\,\Gamma} = \int_{-\infty}^{+\infty} dq^2\,D(q^2,M,\Gamma)\,.
\label{eq:coeff}
\end{equation}
In ZWA, only
the first term of the expansion is taken into account.  The ZWA hence
factorizes the reaction into a production and a decay subprocess, joined 
by an on-shell unstable particle state.

The generalization of the ZWA to unstable particles with nonzero spin
is straightforward, because the pole structure of the propagator is
the same, and the expansion in Eq.~(\ref{BWexp}) is similarly applicable.
Note that in this form the ZWA preserves full spin and polarization 
correlations.  One can take the factorization of production and decay 
subprocesses further and sum/average over the spin/polarization states 
of the intermediate on-shell state in order to fully decouple production 
and decay.  As shown in Ref.\ \cite{Uhlemann:2008pm}, this procedure
is exact for total or sufficiently inclusive cross sections 
of arbitrary resonant processes with an on-shell intermediate state 
decaying via a cubic or quartic vertex.  For less inclusive cross sections,
the decorrelation approximation will cause an additional error.
Note that branching ratios (BR) are defined
and extracted from collider data via
\begin{equation}
\mathrm{BR}_d=\frac{\Gamma_d}{\Gamma}\quad \mathrm{and}\quad 
\sigma = \sigma_p\cdot \mathrm{BR}_d\,,
\label{eq:BR}
\end{equation}
respectively, 
where $\Gamma_d$ ($\Gamma$) is the partial (total) decay width
and $\sigma_p$ is the production cross section of the 
unstable particle in the stable-particle approximation.
The right-hand side of Eq.~(\ref{eq:BR}) implies the application 
of the ZWA and decorrelation approximation.
A process-independent branching ratio definition requires that 
production and decay are fully decoupled.

The uncertainty of the ZWA is typically of ${\cal O}(\Gamma/M)$.  
To be useful for high-energy scattering reactions, the ZWA uncertainty has to 
be comparable to other theoretical and experimental errors.  Since for 
many known and predicted unstable particles $\Gamma/M$ is of ${\cal O}(1\%)$, 
this is indeed the case.
The ZWA is in particular widely used in the analysis of many-particle
signatures due to Beyond-the-Standard-Model (BSM) physics.\footnote{%
A method to include finite-width effects in BSM event 
generators is described in Ref.~\cite{Gigg:2008yc}.}
In this context, long decay chains frequently occur, and a repeated 
application of the ZWA suggests itself.\footnote{%
See Ref.~\cite{Reuter:2012ng} for an example.}  
Since application of 
the ZWA requires the corresponding intermediate unstable particle state 
and restricts it to the mass shell, its application implies that 
subresonant\footnote{%
Subresonant Feynman graphs contain some, but not all unstable states 
that are to be treated in ZWA.} 
and nonresonant amplitude contributions and off-shell effects
are neglected.\footnote{Note that off-shell \underline{contributions} are 
implicitly included.  See Eqs.~(\ref{eq:coeff}), (\ref{eq:ZWA2}) and 
(\ref{eq:ZWA3}).}
For sufficiently inclusive cross sections, these contributions 
are suppressed.
The factorization into production and decay subprocesses 
is also convenient when higher-order corrections are included in 
calculations.\footnote{%
See Ref.~\cite{Falgari:2012sq} for an example.}  
In this case, in ZWA also non-factorizable corrections that connect 
production and decay subprocesses are neglected.\footnote{%
For sufficiently inclusive cross sections, the nonfactorizable corrections 
are of ${\cal O}(\alpha\Gamma/M)$ \cite{Melnikov:1993np,Fadin:1993dz}.}
Calculations for phenomenological studies are hence simplified in two ways: 
First, and most importantly, the number of contributing Feynman graphs is 
significantly reduced as well as the number of kinematic structures that need 
to be included when applying the multichannel Monte Carlo integration 
technique \cite{Berends:1994pv}.  Secondly, the dimensionality of the phase 
space integration is reduced.   
We note that state-of-the-art calculational techniques for 
amplitudes \cite{Binoth:2010ra} and modern computer resources 
allow to include finite-width effects for processes of increasing
complexity, in particular for SM processes and when 
BSM parameter space scans are not carried out.

An alternative interpretation of the ZWA is that 
the leading $q^2$-dependence of the differential cross 
section, i.e.\ the Breit-Wigner resonance, is integrated out.
We illustrate this in Eqs.~(\ref{eq:ZWA1}), (\ref{eq:ZWA2}), and (\ref{eq:ZWA3}),
where $s$ is the total 4-momentum squared, subscripts $p$ and $d$ 
refer to production and decay factors, respectively, and the $M$ and $\Gamma$ 
dependence of $D(q^2,M,\Gamma)$ has been suppressed.
\begin{align}
\label{eq:ZWA1}
\sigma &= \frac{1}{2s}\left[\int_{q^2_\text{min}}^{q^2_\text{max}}
\frac{dq^2}{2\pi}\left(\int d\phi_p|{\cal M}_p(q^2)|^2 
D(q^2)\, \int d\phi_d|{\cal M}_d(q^2)|^2\right)\right]\\
\label{eq:ZWA2}
\sigma_\text{ZWA} &= \frac{1}{2s}\left(\int d\phi_p|{\cal M}_p(M^2)|^2 \right) 
\left(\int_{-\infty}^\infty   \frac{dq^2}{2\pi}\,D(q^2)
\right) \left( \int d\phi_d|{\cal M}_d(M^2)|^2 \right)\\
\label{eq:ZWA3}
\sigma_\text{ZWA} &= \frac{1}{2s}\left(\int d\phi_p|{\cal M}_p(M^2)|^2 \right) 
\frac{1}{2M\Gamma} \left( \int d\phi_d|{\cal M}_d(M^2)|^2 \right)
\end{align}
Evidently, even for a sufficiently small ratio $\Gamma/M$, the ZWA can be 
inadequate if a significant modification of the Breit-Wigner 
lineshape occurs.  Such a modification can be induced by the $q^2$-dependence 
of production and 
decay factors or be due to interference with sub- or nonresonant 
amplitude contributions, i.e.\ corrections or backgrounds which are neglected 
in ZWA.
Here, an important observation is that the Breit-Wigner distribution does 
not drop off nearly as fast as, for instance, a Gaussian. The relative 
contribution of the tail more than $n$ widths from 
the peak can be estimated as $1/(n\pi)$, because \cite{Kauer:2001sp}
\begin{equation}
\label{eq:BWtail}
\int_{(M-n\Gamma)^2}^{(M+n\Gamma)^2}\frac{dq^2}{2\pi}\frac{1}{(q^2-M^2)^2+(M\Gamma)^2}\approx \frac{1}{2M\Gamma}\left(1-\frac{1}{n\pi}\right) .
\end{equation}
Note that the tail region with $|(q^2)^{1/2} - M|>5\,\Gamma$ thus contributes 
more than $6\%$ to the resonant cross section.
Selection cuts or kinematical bounds (e.g.\ $m^2<q^2<s_{\mathrm{max}}$, 
where $m$ is the sum of the masses of the decay products, and $s_{\mathrm{max}}$
is the squared CMS collision energy) can introduce cutoffs, which are not taken 
into account if the ZWA is used.  Eq.~(\ref{eq:BWtail}) can be used to estimate 
the associated error. Typically, kinematical bounds will be far from $M$ 
in terms of $\Gamma$, in which case the uncertainty is negligible.

\begin{figure}[tb]
\centerline{\includegraphics[width=5.8cm, clip=true]{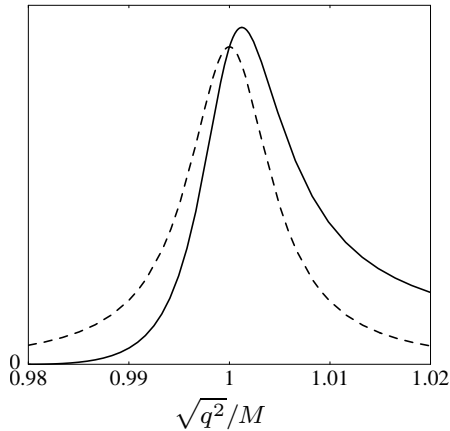}}
\caption{\protect\label{fig:deformation}%
Breit-Wigner lineshape deformation caused by threshold factors when 
a decay daughter mass $m$ approaches the parent mass $M$.
More specifically, 
the Breit-Wigner integrand $D(q^2,M,\Gamma)$ (dashed) 
and the integrand including threshold factors 
$D(q^2,M,\Gamma)\,f(q^2,m^2)/f(M^2,m^2)$ (solid) are shown 
in unspecified normalization 
for a decay via a scalar-fermion-antifermion interaction
as functions of the invariant mass $\sqrt{q^2}$.
$f(x,y) = (x-y)^2/x$, $\Gamma/M=1\%$ and $m=M-2\Gamma$.
Further details can be found in Ref.~\protect\cite{Kauer:2007nt}.}
\end{figure}

Nevertheless, phenomenologically relevant cases where the 
ZWA error exceeds ${\cal O}(\Gamma/M)$ by one order of magnitude 
or more have been identified in the literature 
\cite{Kauer:2012hd,Uhlemann:2008pm,Berdine:2007uv,Kauer:2007zc,Kauer:2007nt}.
A strongly enhanced uncertainty is possible if in ZWA kinematical threshold 
configurations occur.  Consider, for example, a decay chain $A\to B, (C\to D, E)$.
In this case, configurations where $|M_C - (M_A - M_B)|$ or $|M_C - (M_D + M_E)|$ 
is smaller than a few $\Gamma_C$ would be critical.  Such configurations 
are phase space suppressed, and the decays will be rare.
We note that almost degenerate particle masses are not unnatural in SM 
extensions, as evidenced by the SPS benchmark scenarios for SUSY 
searches \cite{Allanach:2002nj}.
The occurring significant deviations between ZWA and off-shell results
can be traced back to threshold factors $\beta(Q_1,Q_2)=(1-Q_1^2/Q_2^2)^{1/2}$,
where $Q_{1,2}$ are (invariant) masses.  The phase-space element contains 
such factors.  For certain interaction types, e.g. scalar-fermion-antifermion, 
additional powers of $\beta$ are contributed by $|{\cal M}|^2$, which enlarges the
phase space region where strongly enhanced deviations occur.  
The effect can be explained by a significant deformation of the Breit-Wigner 
lineshape (peak and tail) caused by powers of the $q^2$-dependent threshold 
factors, as illustrated in Fig.~\ref{fig:deformation}.
The $q^2$-dependence of the residual matrix element 
can cause strongly enhanced ZWA uncertainties even
in the absence of degenerate mass configurations.  
An example is given in Sec.~\ref{lhzwa}.


\section{Zero-Width Approximation Inadequacy for a Light Higgs Boson Signal \label{lhzwa}}


A key objective of current particle physics research is the experimental confirmation of a theoretically consistent description of elementary particle masses.  
In the SM, this is 
achieved through the Higgs mechanism \cite{%
Higgs:1964ia,Higgs:1964pj,Higgs:1966ev,Englert:1964et,Guralnik:1964eu}, 
which predicts the existence of one physical Higgs boson.
A thorough examination of the recently discovered candidate 
SM Higgs boson \cite{Aad:2012tfa,Chatrchyan:2012ufa} with 
$M_H\approx 125$ GeV is in progress, 
and its properties are in agreement with theoretical expectations.
No compelling deviations have been observed so far.
In this situation, it is prudent to examine the accuracy of 
theoretical predictions for light Higgs production and decay 
that are used in experimental analyses.
For light Higgs masses, 
the loop-induced gluon-fusion production ($gg\to H$) 
dominates \cite{Georgi:1977gs}.  Next-to-leading order QCD corrections 
have been calculated in the heavy-top limit \cite{Dawson:1990zj} and  
with finite $t$ and $b$ mass effects \cite{%
Djouadi:1991tka,Graudenz:1992pv,Spira:1995rr}, and were found to be as large as
80--100\% at the Large Hadron Collider (LHC).  
This motivated the calculation of next-to-next-to-leading 
order QCD corrections \cite{Harlander:2002wh,Anastasiou:2002yz,Ravindran:2003um} 
enhanced by soft-gluon resummation at next-to-next-to-leading logarithmic 
level \cite{Catani:2003zt,deFlorian:2011xf} and beyond \cite{Moch:2005ky}.
In addition to higher-order QCD corrections, electroweak corrections 
have been computed and found to be at the 1--5\% 
level \cite{Djouadi:1994ge,Actis:2008ug,Anastasiou:2008tj}.
Further references on all aspects of Higgs physics at the LHC can be found 
in Refs.~\cite{Dittmaier:2011ti,Dittmaier:2012vm}.

A comparison of the ZWA and finite-width Higgs propagator schemes in 
inclusive Higgs production and decay in gluon fusion was carried out in 
Refs.~\cite{Anastasiou:2011pi,Anastasiou:2012hx} for Higgs masses between 
120 and 800 GeV.\footnote{%
We note that the pinch-technique approach yields a theoretically 
well-behaved Dyson-resummed Higgs boson propagator 
\protect\cite{Papavassiliou:1997pb,Papavassiliou:1997fn}, 
which does not suffer from unphysical absorptive effects, as seen 
in the comparison in Fig.~5 of Ref.\ \cite{Papavassiliou:1997pb}.}
Overall, good agreement with the expected uncertainty of 
${\cal O}(\Gamma_H/M_H)$ was found.  In particular, for light Higgs 
masses ($M_H < 300$ GeV) a relatively small error of ${\cal O}(1\%)$ was 
found \cite{Anastasiou:2011pi}, leading to the conclusion that the 
ZWA should be an adequate treatment for a light Higgs boson where the
Higgs width is very small compared to its mass \cite{Anastasiou:2012hx}.  
Curiously, a closer inspection of the results for $M_H=120$ GeV reveal 
that the deviation between ZWA and fixed-width Breit-Wigner scheme 
($0.5\%$) is two orders of magnitude larger than 
$\Gamma_H/M_H$ ($4\cdot 10^{-5})$.  Based on the discussion in 
Sec.~\ref{sec:zwa}, it is suggestive to interpret this as evidence 
for a significant deformation of the Breit-Wigner lineshape for a 
light Higgs boson.  Such deformations were first predicted and 
thoroughly studied in Ref.~\cite{Kauer:2012hd}.  They can be 
traced back to the dependence of the Higgs decay amplitude ${\cal M}_d$ on 
the Higgs virtuality $q^2$ for different decay modes 
(cf.\ Eqs.~(\ref{eq:ZWA1}) and (\ref{eq:ZWA3})).  One has, for instance,\footnote{Here, $\gtrsim$ implies above, but not too far above threshold.}
\begin{align}
&|{\cal M}_d(H\to f\bar{f})|^2\;\ \sim\ M_f^2 q^2\quad \mathrm{for}\ \sqrt{q^2}\gtrsim 2\,M_f\,,\\
&|{\cal M}_d(H\to VV)|^2\!\ \sim\ (q^2)^2\;\quad \mathrm{for}\ \sqrt{q^2}\gtrsim 2\,M_V\,,
\end{align}
for Higgs boson decays to fermions $f$ or weak bosons $V$.  
For the $H\to WW$ and $H\to ZZ$ decay modes of a light Higgs 
boson with resonance below the $VV$ threshold, a remarkable effect 
occurs above the $VV$ threshold (far away from the resonance peak): 
the leading $(q^2)^{-2}$ dependence of the off-shell squared Higgs 
propagator $|D|^2$ and the leading $(q^2)^{2}$ dependence of 
$|{\cal M}_d|^2$ largely compensate.  The Higgs lineshape is therefore 
strongly enhanced for 
$(q^2)^{1/2}> 2\,M_V$ compared to the Breit-Wigner expectation, which is 
given by
\begin{equation}
\label{eq:ZWA_MVV}
\left(\frac{d\sigma}{dM_{VV}}\right)_{\!\text{ZWA}} =\ \sigma_{H,\text{ZWA}}\;\frac{M_H\Gamma_H}{\pi}\;\frac{2M_{VV}}{\left(M_{VV}^2-M_H^2\right)^2+(M_H\Gamma_H)^2} \;.
\end{equation}
\begin{figure}[tb]
\centerline{\includegraphics[width=3.6in, clip=true]{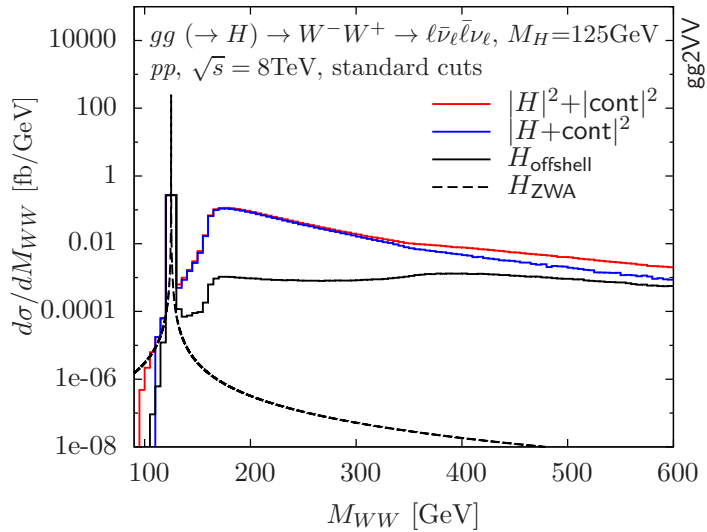}}
\caption{\protect\label{fig:lvlv_MWW_l}
$M_{WW}$ distributions for 
$gg\ (\to H)\to W^-W^+\to \ell\bar{\nu}_\ell \bar{\ell}\nu_\ell$ 
in $pp$ collisions at $\sqrt{s} = 8$\,TeV 
for $M_H=125$\,GeV and $\Gamma_H = 0.004434$\,GeV 
calculated at LO with \textsf{gg2VV} \protect\cite{gg2VV}.
The ZWA distribution (black, dashed) as defined in Eq.~(\ref{eq:ZWA_MVV})
in the main text, the off-shell 
Higgs distribution (black, solid), the 
$d\sigma(|{\cal M}_H + {\cal M}_\text{cont}|^2)/dM_{WW}$ 
distribution (blue) and the 
$d\sigma(|{\cal M}_H|^2 + |{\cal M}_\text{cont}|^2)/dM_{WW}$ 
distribution (red) are shown.
Standard cuts are applied: 
$p_{T\ell} > 20$\,GeV, $|\eta_\ell| < 2.5$, $\sla{p}_T > 30$\,GeV, 
$M_{\ell\ell} >$ 12\,GeV.
Differential cross sections for a single lepton flavor combination are displayed. 
No flavor summation is carried out for charged leptons or neutrinos.
Further details can be found in Ref.~\protect\cite{Kauer:2012hd}.}
\end{figure}
\begin{figure}[tb]
\centerline{\includegraphics[width=3.6in, clip=true]{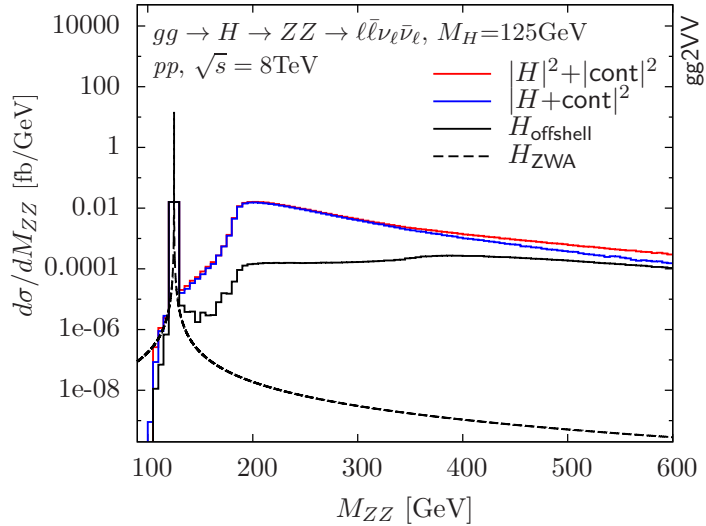}}
\caption{\protect\label{fig:2l2v_125_MZZ_l}
$M_{ZZ}$ distributions for $gg\ (\to H)\to ZZ\to \ell\bar{\ell}\nu_\ell\bar{\nu}_\ell$ 
for $M_H=125$\,GeV.
Applied cuts: $p_{T\ell} > 20$\,GeV, $|\eta_\ell| < 2.5$, 
$76$\,GeV $< M_{\ell\ell} < 106$\,GeV,
$\sla{p}_T > 10$\,GeV.
Other details as in Fig.~\protect\ref{fig:lvlv_MWW_l}.}
\end{figure}
The Breit-Wigner expectation ($H_{\mathrm{ZWA}}$) and the enhanced off-shell  
distribution ($H_{\mathrm{offshell}}$) are illustrated in 
Figs.~\ref{fig:lvlv_MWW_l} and \ref{fig:2l2v_125_MZZ_l} using the 
$gg\to H\to W^-W^+\to \ell\bar{\nu}_\ell \bar{\ell}\nu_\ell$ and 
$gg\to H\to ZZ\to \ell\bar{\ell}\nu_\ell\bar{\nu}_\ell$ processes, respectively.
The differential 
cross section in the plateau-like finite-width tail is approximately two to 
three orders of 
magnitude smaller than in the resonance region.  However, the plateau extends from 
the $VV$ threshold to beyond 600 GeV.  The integrated cross section in this region 
far from resonance 
thus contributes ${\cal O}(10\%)$ to the total finite-width cross section, more 
specifically, $16\%$ and $37\%$ in Figs.~\ref{fig:lvlv_MWW_l} and 
\ref{fig:2l2v_125_MZZ_l}, respectively.  ZWA errors of ${\cal O}(10\%)$ can 
therefore occur in $H\to VV$ decay modes, despite 
$\Gamma_H/M_H < 10^{-4}$.\footnote{%
For $gg \to H \to \gamma\gamma$ the effect
is drastically reduced and confined to the region $M_{\gamma\gamma}$ between 157 GeV 
and 168 GeV, where the distribution is already five orders of magnitude smaller 
than in the resonance region.}
We emphasize that $H\to VV$ modes in Higgs production 
channels other than gluon fusion also exhibit an enhanced off-shell tail, 
since the effect is caused by the decay amplitude.

Evidently, the ZWA caveat also applies 
to Monte Carlo generators that approximate off-shell effects with an 
ad hoc Breit-Wigner reweighting of the on-shell propagator  
(cf.\ Eq.\ (\ref{eq:ZWA_MVV})).
Furthermore, the ZWA limitations are relevant for the extraction of Higgs 
couplings, which is initially being performed using the ZWA.  
The findings of Ref.~\cite{Kauer:2012hd} make clear that off-shell 
effects have to be included in future Higgs couplings analyses.

\begin{figure}[tb]
\vspace*{0.4cm}
\centerline{%
\hspace*{0.35cm}\includegraphics[height=1.95cm, clip=true]{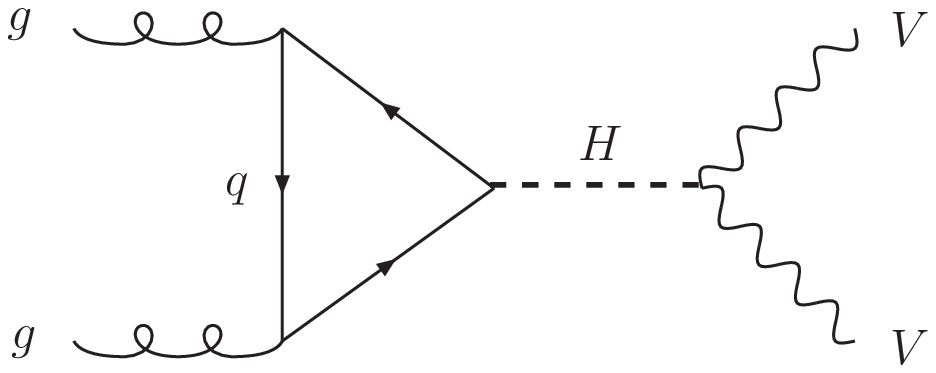}
\includegraphics[height=1.95cm, clip=true]{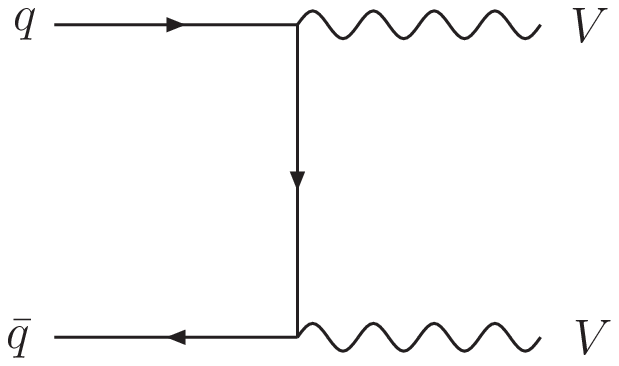}
\includegraphics[height=1.95cm, clip=true]{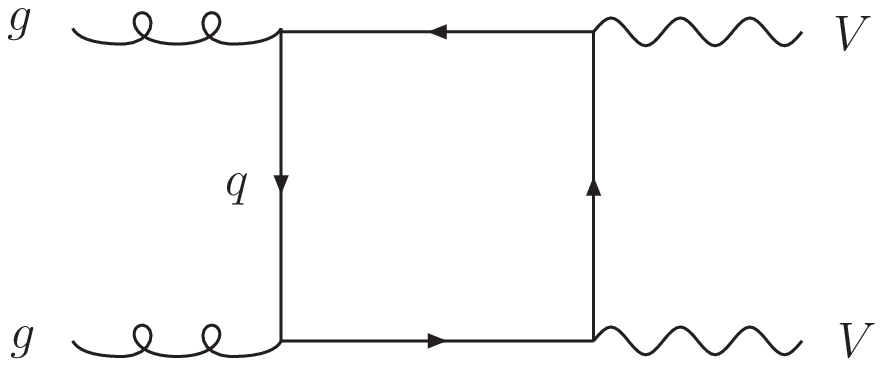}
}
\caption{\protect\label{fig:graphs}%
Representative Feynman graphs for the Higgs signal process (left) and the $q\bar{q}$- (center) and $gg$-initiated (right) continuum background processes.}
\end{figure}

Above the $VV$ threshold, the $gg\to VV$ continuum background is 
large and sizable signal-background interference (see Fig.~\ref{fig:graphs}, 
left and right) occurs.  Resonance-continuum interference in 
$gg\ (\to H)\to VV$
has been studied in Refs.\ \cite{Kauer:2012hd,Glover:1988fe,Glover:1988rg,Seymour:1995qg,Binoth:2006mf,Accomando:2007xc,Campbell:2011cu,Kauer:2012ma,Passarino:2012ri,Campanario:2012bh,Bonvini:2013jha} 
and for related processes in Refs.\ \cite{Dixon:2003yb,Dixon:2008xc,Accomando:2011eu,Martin:2012xc,deFlorian:2013psa,Martin:2013ula,Accomando:2013sfa}.\footnote{For studies of the $q\bar{q}$ and $gg$ continuum background (see Fig.~\ref{fig:graphs}, center and right), we refer the reader to Refs.\ \protect\cite{Campbell:1999ah,Binoth:2005ua,Binoth:2008pr,Campbell:2011bn,Melia:2011tj,Frederix:2011ss,Melia:2012zg,Agrawal:2012df} and references therein.}
Due to the enhanced Higgs cross section above the $VV$ threshold, integrated cross 
sections can be affected by ${\cal O}(10\%)$ signal-background interference effects, 
which are hence also displayed in Figs.~\ref{fig:lvlv_MWW_l} and 
\ref{fig:2l2v_125_MZZ_l}.

\begin{figure}[tb]
\centerline{\includegraphics[width=3.6in, clip=true]{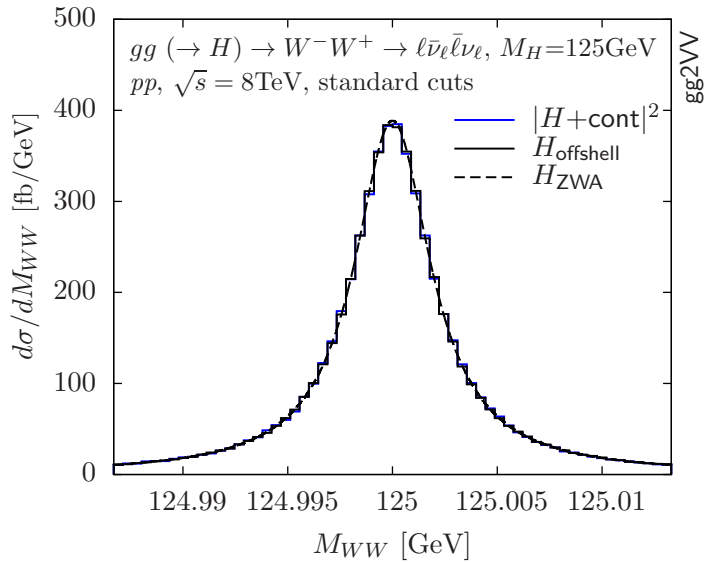}}
\caption{\protect\label{fig:lvlv_MWW_s}
$M_{WW}$ distributions for 
$gg\ (\to H)\to W^-W^+\to \ell\bar{\nu}_\ell \bar{\ell}\nu_\ell$ 
for $M_H=125$\,GeV.  Off-shell and interference effects in the vicinity of the
Higgs resonance are shown.
Other details as in Fig.~\protect\ref{fig:lvlv_MWW_l}.}
\end{figure}

In the vicinity of the 
Higgs resonance, finite-width and Higgs-continuum interference effects are 
negligible for $gg\ (\to H)\to VV$ 
if $M_H \ll 2M_{V}$, as shown in Fig.~\ref{fig:lvlv_MWW_s} for 
$gg\ (\to H)\to W^-W^+\to \ell\bar{\nu}_\ell \bar{\ell}\nu_\ell$.
For weak boson decays that permit the reconstruction 
of the Higgs invariant mass, the experimental procedure focuses on the 
Higgs resonance region and for $M_H \ll 2M_{V}$ the 
enhanced off-shell region is thus typically excluded.

\begin{figure}[tb]
\centerline{\includegraphics[width=3.6in, clip=true]{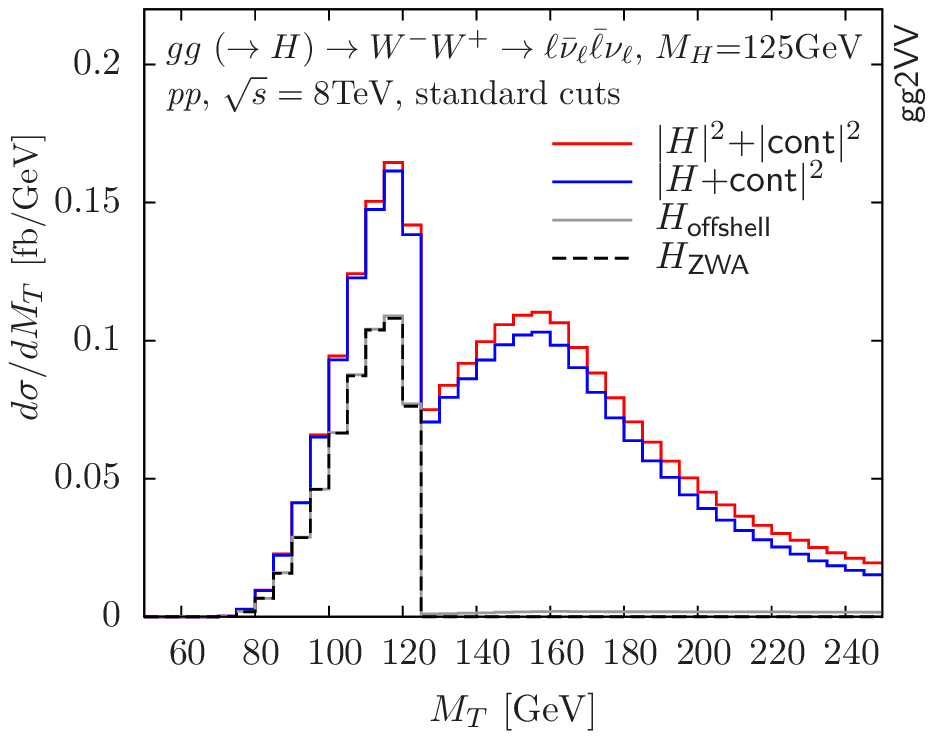}}
\caption{\protect\label{fig:lvlv_MT1_m}
Transverse mass distributions for 
$gg\ (\to H)\to W^-W^+\to \ell\bar{\nu}_\ell \bar{\ell}\nu_\ell$ 
and $M_H=125$ GeV.
Off-shell and interference effects in the region of the
Higgs resonance and the $W$-pair threshold are shown.
$M_{T}$ is defined in Eq.\ (\ref{eq:MT1}) in the main text.
Other details as in Fig.\ \protect\ref{fig:lvlv_MWW_l}.}
\end{figure}

For $H\to VV$ channels that do not allow to reconstruct the Higgs 
invariant mass, the tail contribution can nevertheless be reduced 
significantly by means of optimized selection cuts.  In Table \ref{tab:lvlv}, 
we demonstrate this for 
$gg\ (\to H)\to W^-W^+\to \ell\bar{\nu}_\ell \bar{\ell}\nu_\ell$.
\begin{table}[tb]
\caption{\protect\label{tab:lvlv}
Cross sections for 
$gg\ (\to H)\to W^-W^+\to \ell\bar{\nu}_\ell \bar{\ell}\nu_\ell$ 
and $M_H=125$\,GeV with standard cuts, Higgs search cuts and additional cut on the 
transverse mass $M_T$ defined in Eq.\ (\protect\ref{eq:MT1}) in the main text.
Standard cuts: as in Fig.~\protect\ref{fig:lvlv_MWW_l}.  
Higgs search cuts: standard cuts and 
$M_{\ell\ell} <$ 50\,GeV, $\Delta\phi_{\ell\ell} < 1.8$.
The zero-width approximation (ZWA) and off-shell 
Higgs cross sections, the $gg$ continuum cross section and the sum of
off-shell Higgs and continuum cross sections including interference are given.  
The accuracy of the ZWA and the impact of off-shell 
effects are assessed with 
$R=\sigma_{H,\text{ZWA}}/\sigma_{H,\text{offshell}}$.
The integration error is given in brackets.
Other details as in Fig.~\protect\ref{fig:lvlv_MWW_l}.}
\begin{center}
\begin{tabular}{@{}lccccc@{}}
\hline
\multicolumn{6}{c}{$gg\ (\to H)\to W^-W^+\to \ell\bar{\nu}_\ell \bar{\ell}\nu_\ell$,\ \;$\sigma$ [fb],\ \;$pp$,\ \;$\sqrt{s} = 8$ TeV,\ \;$M_H=125$ GeV} \\
\hline
selection cuts & $H_\text{ZWA}$ & $H_\text{offshell}$ & cont & $|H_\text{ofs}$+cont$|^2$ & $R$ \\
\hline
standard cuts & 2.707(3)\phantom{0} & 3.225(3) & 10.493(5) & 12.241(8) & 0.839(2)\phantom{0} \\
Higgs search cuts & 1.950(1)\phantom{0} & 1.980(1) & \phantom{0}2.705(2) & \phantom{0}4.497(3) & 0.9850(7) \\
\hline
$0.75M_H < M_T < M_H$ & 1.7726(9) & 1.779(1) & \phantom{0}0.644(1) & \phantom{0}2.383(2) & 0.9966(8) \\
\hline
\end{tabular}
\end{center}
\end{table}
Here, the search selection has additional cuts, in particular an upper 
bound on the invariant mass of the observed dilepton system, which 
significantly reduce the contribution from the region with $M_{WW} \gg 2M_W$.  
The result is a substantial mitigation of the off-shell (see Table \ref{tab:lvlv}) 
and interference effects (see Ref.~\cite{Kauer:2012hd}).
As first noted in Ref.\ \cite{Campbell:2011cu}, a very effective means to 
suppress the tail contribution is provided by cuts on transverse mass observables 
 \cite{Barr:2009mx}, 
which are designed to have the physical mass of the decaying parent particle 
(the invariant mass in the off-shell case) as upper bound.
For the process considered here, the best-performing transverse mass is defined by
\begin{equation}
\label{eq:MT1}
M_T=\sqrt{(M_{T,\ell\ell}+\sla{p}_{T})^2-({\bf{p}}_{T,\ell\ell}+{\sla{\bf{p}}}_T)^2}\ \;\mathrm{with}\ \;M_{T,\ell\ell}=\sqrt{p_{T,\ell\ell}^2+M_{\ell\ell}^2}\,.
\end{equation}
$M_T$ distributions are shown in Fig.\ \ref{fig:lvlv_MT1_m}, which demonstrates 
that a $M_T < M_H$ cut strongly suppresses off-shell 
as well as interference effects.
Table \ref{tab:lvlv} shows that the application of this 
$M_T$ cut reduces the ZWA error to the sub-percent level.
Note, however, that the cut on $M_T$ cannot completely eliminate 
the unwanted $M_{VV}>2M_V$ contribution: $M_T \leq M_{VV}$
for all phase space configurations, and a small contamination remains.
The efficiency of a transverse mass cut 
for $gg\ (\to H)\to ZZ \to \ell\bar{\ell}\nu_\ell\bar{\nu}_\ell$ 
is illustrated in Table \ref{tab:2l2v_125}.
\begin{table}[tb]
\caption{\protect\label{tab:2l2v_125}
Cross sections for $gg\ (\to H)\to ZZ \to \ell\bar{\ell}\nu_\ell\bar{\nu}_\ell$ and $M_H=125$\,GeV without and with transverse mass cut.
Applied cuts: $p_{T\ell} > 20$ GeV, $|\eta_\ell| < 2.5$, 
$76$ GeV $< M_{\ell\ell} < 106$ GeV,
$\sla{p}_T > 10$ GeV.
$M_{T}$ is defined in Eq.\ (\ref{eq:MT1}) in the main text.
Other details as in Table \protect\ref{tab:lvlv}.}
\begin{center}
\begin{tabular}{@{}lccccc@{}}
\hline
\multicolumn{6}{c}{$gg\ (\to H)\to ZZ \to \ell\bar{\ell}\nu_\ell\bar{\nu}_\ell$,\ \;$\sigma$ [fb],\ \;$pp$,\ \;$\sqrt{s} = 8$ TeV,\ \;$M_H=125$ GeV} \\
\hline
$M_T$ cut & $H_\text{ZWA}$ & $H_\text{offshell}$ & cont & $|H_\text{ofs}$+cont$|^2$ & $R$ \\
\hline
none & 0.1593(2) & 0.2571(2) & 1.5631(7) & 1.6376(9) & 0.6196(7) \\
$M_{T} < M_H$ & 0.1593(2) & 0.1625(2) & 0.4197(5) & 0.5663(6) & 0.980(2)\phantom{0} \\
\hline
\end{tabular}
\end{center}
\end{table}


\section{Summary \label{summary}}

The evolution of methods for perturbative calculations 
that involve unstable particles has been reviewed.  A  
general formalism that has a rigorous field-theoretical 
foundation and provides a practicable and efficient 
implementation which returns reliable results of the 
desired precision for all phenomenologically relevant 
observables is not yet known.  This fact was used to 
motivate a review of on-mass-shell approximations.
The zero-width approximation, a.k.a.\ narrow-width approximation,
restricts the intermediate unstable particle state 
to the mass shell and, when combined with the decorrelation 
approximation, fully factorizes the production and decay 
subprocesses.  Both approximations are implicitly applied 
when extracting branching ratios from collider data.  
The ZWA strongly reduces the complexity of 
calculations of cross sections for many-particle processes in 
Standard Model extensions and/or higher-order corrections.
The uncertainty of the ZWA is typically of ${\cal O}(\Gamma/M)$, 
where $M$ ($\Gamma$) is the mass (width) of the unstable 
particle, but can be much larger in special cases 
where a significant 
modification of the Breit-Wigner lineshape occurs.  Such a 
modification can be induced by the $q^2$-dependence of production 
and decay factors, where $q^2$ is the unstable particle virtuality, 
or be due to interference with sub- or nonresonant 
amplitude contributions, i.e.\ corrections or backgrounds which are 
neglected in ZWA.  A strongly enhanced error is, for instance, 
possible if in ZWA kinematical threshold configurations occur.  

A thorough examination of the recently discovered candidate 
SM Higgs boson is in progress, and its properties are in good 
agreement with theoretical expectations.
It is thus prudent to examine the accuracy of 
theoretical predictions for light Higgs production and decay 
that are used in experimental analyses.
For the SM Higgs boson with $M_H\approx 125$\ GeV, one has
$\Gamma_H/M_H < 10^{-4}$, which suggests an excellent accuracy 
of the ZWA.  We have demonstrated that the ZWA is 
in general not adequate and the error estimate ${\cal O}(\Gamma_H/M_H)$ 
is not reliable for a light Higgs boson.
The inclusion of off-shell contributions is essential to obtain 
an accurate Higgs signal normalization at the $1\%$ precision level
as well as correct kinematic distributions.  ZWA deviations are 
particularly large for $gg\to H\to VV$ processes ($V=W,Z$).
To be more specific, without optimized selection cuts they are of 
\mbox{${\cal O}(5\,$--$\,10\%)$}.  
The ZWA caveat also applies to Monte Carlo generators that approximate 
off-shell effects with an ad hoc Breit-Wigner reweighting of the on-shell 
propagator.  Moreover, the ZWA limitations are relevant for the extraction of 
Higgs couplings, which is initially being performed using the ZWA.  
Our findings make clear that off-shell effects have to be included in 
future Higgs couplings analyses.
The unexpected off-shell effect can be traced back to the dependence of 
the decay amplitude on the Higgs virtuality $q^2$.  For $H\to VV$ 
decay modes one finds that the $(q^2)^2$ dependence of the squared 
modulus of the decay amplitude above the $VV$ threshold compensates the 
$q^2$-dependence of the Higgs propagator, which 
causes a strongly enhanced off-shell cross section 
up to invariant masses of about $600$~GeV.
We note that $H\to VV$ modes in Higgs production 
channels other than gluon fusion also exhibit an enhanced off-shell tail, 
since the effect is caused by the decay amplitude.
The total $gg\to H \to VV$ cross section receives an 
${\cal O}(10\%)$ off-shell correction and a similar correction 
due to signal-background interference.  These effects are negligible 
in the vicinity of the Higgs resonance if $M_H \ll 2M_{V}$.  
If the Higgs mass can be reconstructed exclusion of the problematic 
region above the $VV$ threshold is straightforward.
For $H\to VV$ channels where this is not the case, 
the enhanced tail can be strongly suppressed  
by applying a $M_T<M_{H}$ cut on a suitable transverse mass 
observable $M_T$.


\section*{Acknowledgments}

Important contributions of G.~Passarino and 
financial support from HEFCE, STFC and the IPPP Durham are gratefully 
acknowledged.

\end{document}